\documentclass[apjl]{emulateapj}

\usepackage{epsfig,graphicx}
\usepackage{txfonts}
\usepackage{natbib}

\newcommand{\nh}{N$_2$H$^+$}
\newcommand{\ch}{CH$_3$CN}
\newcommand{\kms}{km~s$^{-1}$}
\newcommand{\msol}{M$_{\odot}$}

\slugcomment{To appear in The Astrophysical Journal Letters}

\shorttitle{Convergent flows and low-velocity shocks in DR21(OH)}
\shortauthors{T. Csengeri et al.}

\begin{document}


\title{Convergent flows and low-velocity shocks in DR21(OH)}


\author{T. Csengeri}
\affil{Max Planck Institute for Radioastronomy, Auf dem H\"ugel 69, 53121 Bonn, Germany} \email{ctimea@mpifr-bonn.mpg.de}
\affil{Laboratoire AIM Paris Saclay, CEA-INSU/CNRS-Universit\'e Paris Diderot, IRFU/SAp CEA-Saclay, 91191 Gif-sur-Yvette, France}
\author{S. Bontemps}
\affil{OASU/LAB-UMR5804, CNRS, Universit\'e Bordeaux 1, 33270 Floirac, France }   
\author{N. Schneider and F. Motte}
\affil{Laboratoire AIM Paris Saclay, CEA-INSU/CNRS-Universit\'e Paris Diderot, IRFU/SAp CEA-Saclay, 91191 Gif-sur-Yvette, France}
\author {F. Gueth}
\affil{IRAM, 300 rue de la piscine, 38406, Saint Martin d'H{\`e}res, France}
\author {J. L. Hora}
\affil{Harvard-Smithsonian Center for Astrophysics, 60 Garden Street, MS-65, Cambridge, MA 02138}


\begin{abstract}
DR21(OH) is a pc-scale massive, $\sim$7000~M$_{\odot}$ clump hosting three massive dense cores (MDCs) at an early stage of their evolution. We present a high angular-resolution mosaic, covering $\sim$70$\arcsec\times$100\arcsec, with the IRAM PdBI at 3~mm to trace the dust continuum emission and the {\nh} (J=1--0) and {\ch} (J=5--4) molecular emission. The cold, dense gas traced by the 
compact emission in {\nh} is associated with the three MDCs and shows several velocity components towards each MDC. 
These velocity components reveal local shears in the velocity fields which are best interpreted as convergent flows. Moreover, we report the detection of weak extended emission from {\ch} at the position of the {\nh} velocity shears.  We propose that this extended {\ch} emission is tracing warm gas associated with the low-velocity shocks expected at the location of convergence of the flows where velocity shears are observed. This is the first detection of low-velocity shocks associated with small (sub-parsec) scale convergent flows which are proposed to be at the origin of the densest structures and of the formation of (high-mass) stars. In addition, we propose that MDCs may be active sites of star-formation for more than a crossing time as they continuously receive material from larger scale flows as suggested by the global picture of dynamical, gravity driven evolution of massive clumps which is favored by the present observations.

\end{abstract}


\keywords{ISM: kinematics and dynamics --- stars: formation }



\section{Introduction}

Two competing scenarios are challenged by observations to describe the formation of rich clusters and of high-mass stars: a quasi-static evolution (also known as core accretion model) versus a highly dynamical model. The first one is a turbulence regulated scenario, where a high level of micro-turbulence acts as an effective additional thermal pressure to balance gravity (e.g.\,\citealp{MT02}). In the second one, massive cores form and evolve via highly dynamical processes (e.g.\,  \citealp{BB06}; \citealp{VS07}; \citealp{Heitsch2008}; \citealp{H08}; \citealp{KH09}), where convergent flows driven by large-scale turbulence, gravity and Galactic motions create dense structures down to small scales by shock-dissipation at their stagnation points (e.g.\,\citealp{Field08}; \citealp{VS11}).

In the \object{Cygnus X} region a population of massive dense cores (hereafter MDCs) was revealed by a systematic dust continuum survey \citep{M07}. The most massive of them are the birth place of high-mass stars~\citep{B09}. Among the MDCs more massive than 40~{\msol} all show star formation activity, leading to the suggestion that the formation of high-mass stars within MDCs is a fast process \citep{M07}. \citet{Csengeri10} indeed recognized the important role of dynamical processes inside the young IR-quiet MDCs with crossing times only slightly larger than the local free-fall times, confirming a fast evolution. The precise origin of the MDCs and of their prime properties (mass, size, spatial distribution) is still uncertain as well as one needs to understand how a rich cluster can be formed from relatively small MDCs (see discussion in Bontemps et al. 2010). \citet{Csengeri10} also proposed that the massive protostars in the MDCs are found at the location of velocity shears by the convergence of small scale flows. 
The existence of these convergent flows needs however confirmation by, for instance, the direct detection of the associated low-velocity shocks~(e.g.\,\citealp{Klessen}). 

\begin{figure*}[!htpb]
   \centering
   \includegraphics[height=4.32cm,angle=-90]{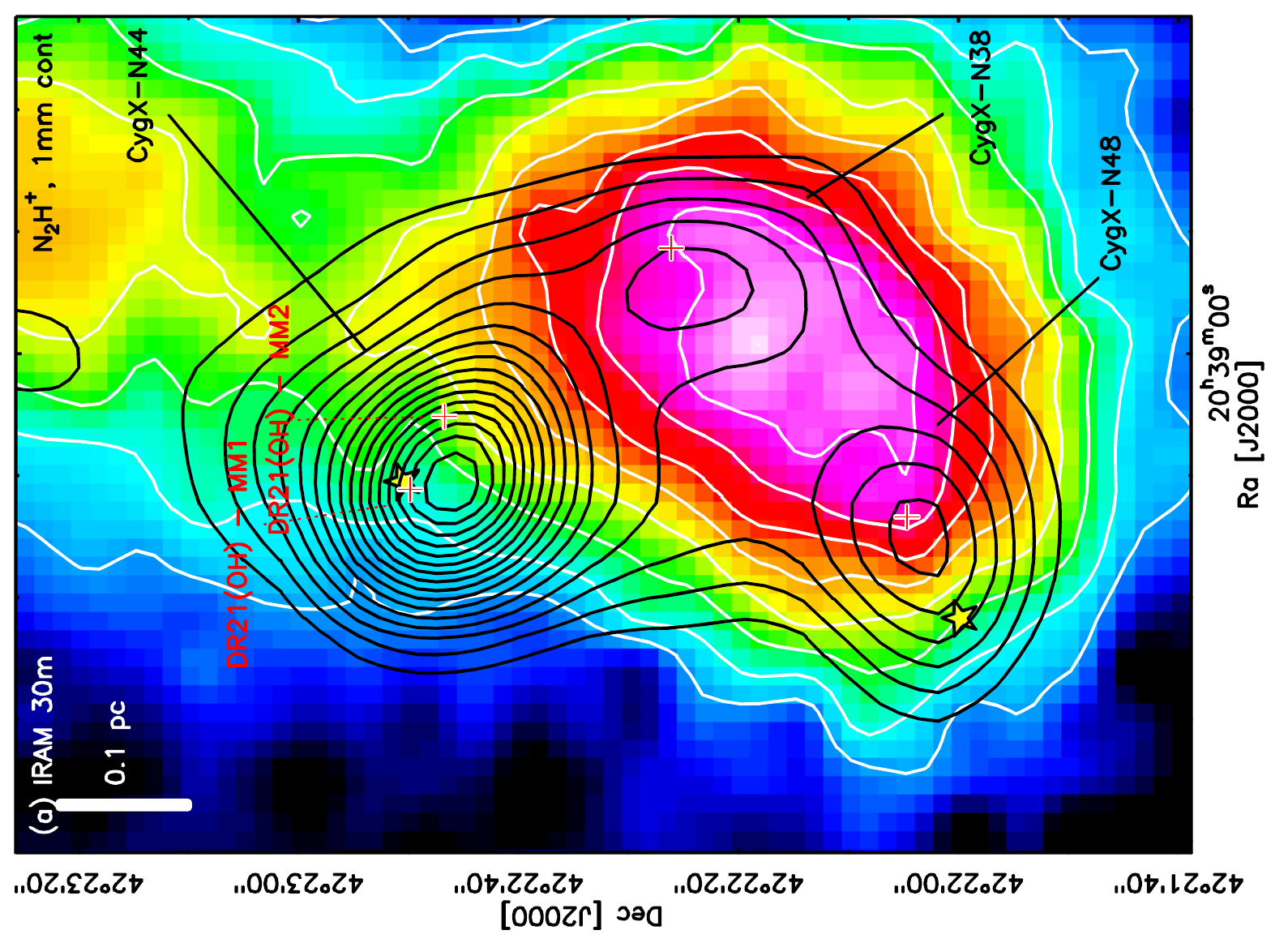}
   \includegraphics[height=3.95cm,angle=-90]{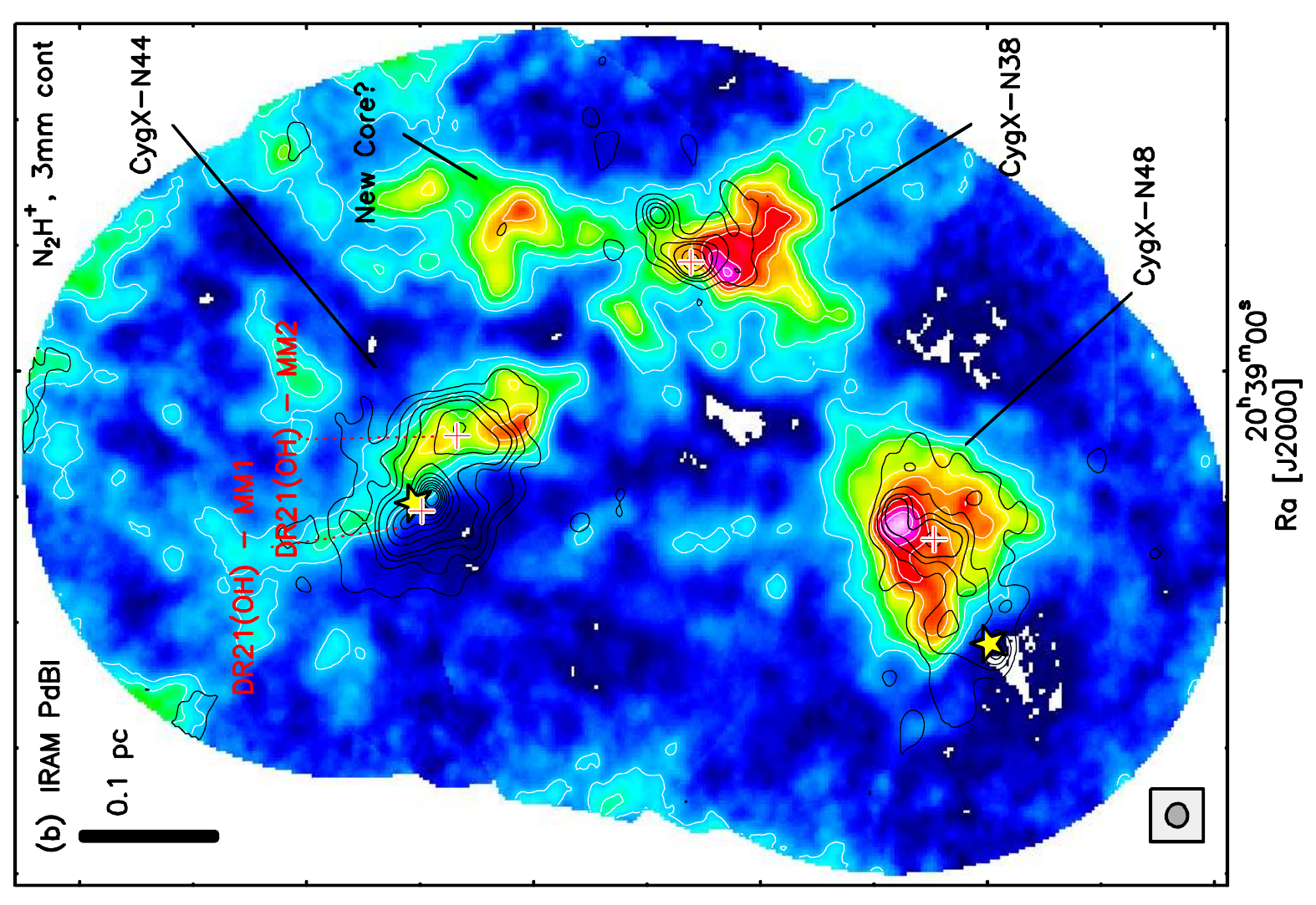}
   \includegraphics[height=3.95cm,angle=-90]{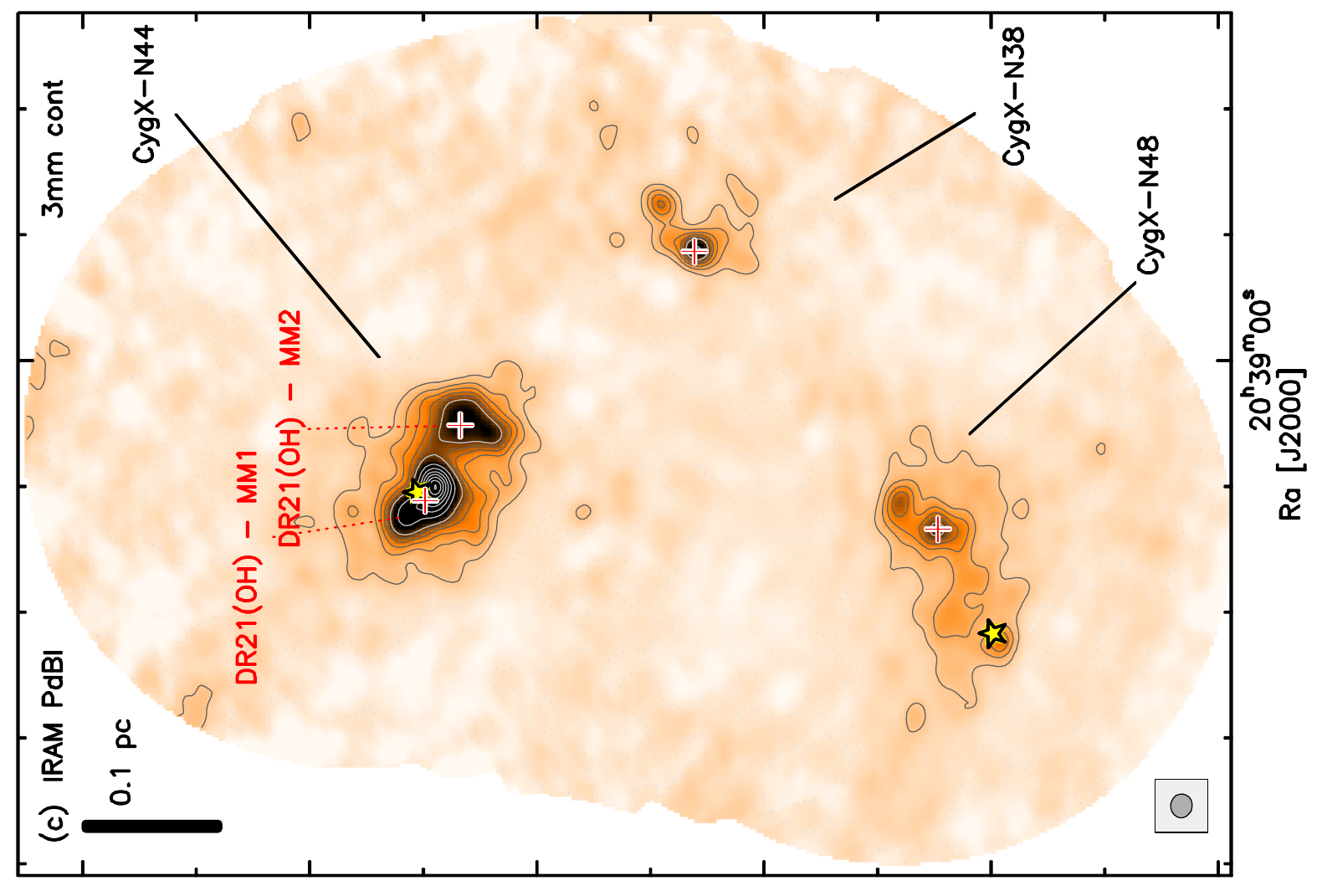}
   \includegraphics[height=3.95cm,angle=-90]{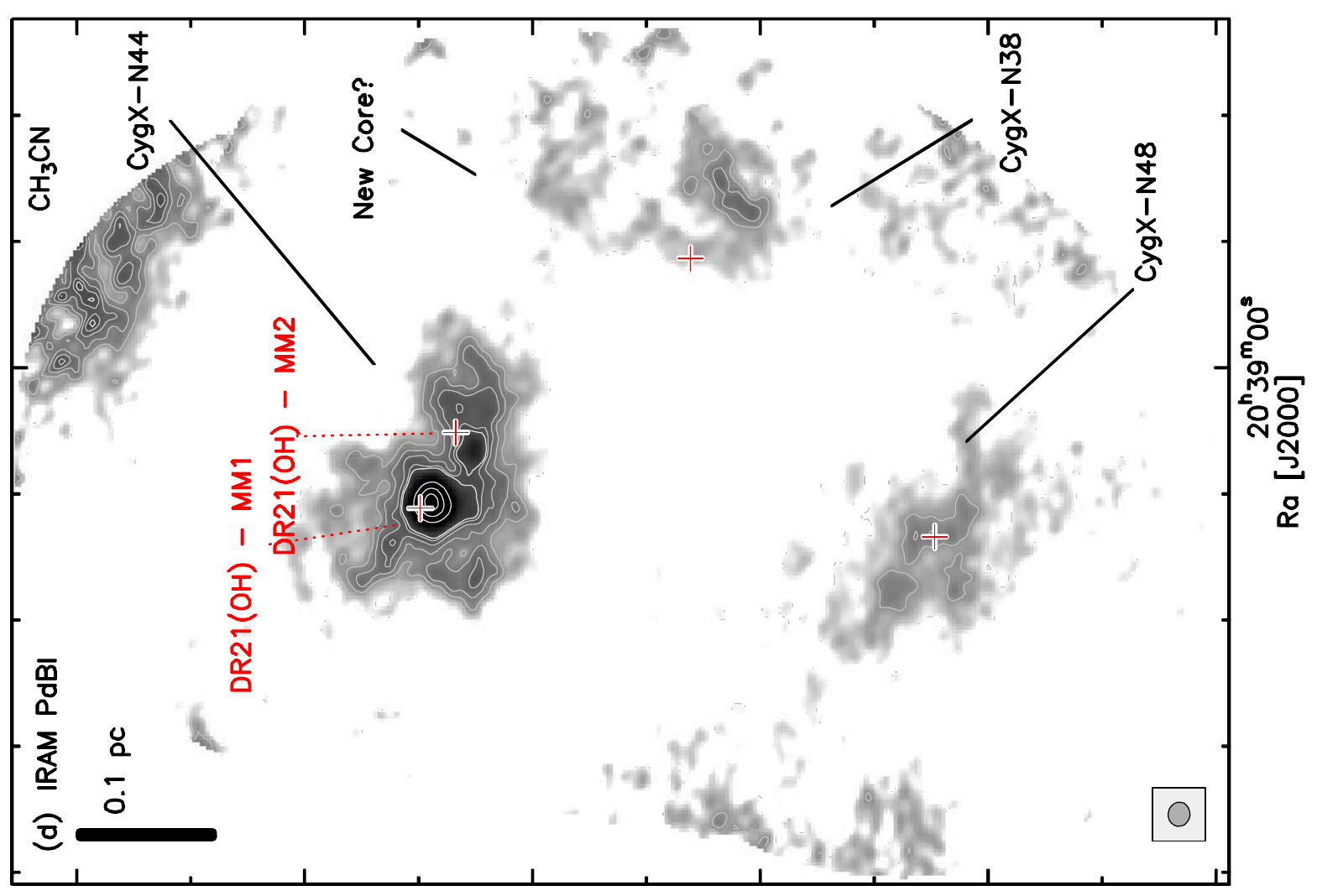}

      \caption{ {\bf a)} Color scale shows the integrated intensity map of the {\nh} (J=1--0) line obtained by \citet{Schneider_prep} with the IRAM 30m telescope (integration range is between $-10$ to $+0.5$~{\kms}). White contours indicate the 20\% level of the peak increasing by 10\%. Black contours show 1.2 mm emission obtained by the MAMBO-2 survey of Cygnus-X by \citet{M07}. Stars indicate sources with compact free-free emission (VLA archival data). Red crosses show DR21(OH)-MM1, -MM2 and the peak 3mm PdBI continuum emission towards CygX-N38 and -N48, respectively.  Labels mark the MDCs from \citet{M07}.  {\bf b)} Map of the same line obtained with the PdBI (without zero-spacings) integrated over the more isolated, F=101-012 hyperfine component corresponding to $-10$ to $+5$~{\kms}, white contours indicate the 20\% level of the peak increasing by 10\%. Black contours show the 3mm dust continuum. {\bf c)} Map of the 3mm dust continuum emission. Gray contours go from 5$\sigma$ to 30$\sigma$ by 5$\sigma$ ($\sigma=0.3$~mJy/beam), light grey contours go from 30$\sigma$ to 150$\sigma$ by 15$\sigma$. {\bf d)} Map of integrated intensity of {\ch} emission over all K-components from $-10$ to $+55$~{\kms}. Gray contours go from 3$\sigma$ to 20$\sigma$ by 3$\sigma$, then by 40$\sigma$ ($\sigma=35$~mJy/beam). (For a color figure see the online edition.) 
      }
         \label{fig:intro_n2h}
   \end{figure*}

The clump associated with \object{DR21(OH)} is the most massive 1~pc-scale clump of the whole Cygnus X region. It is embedded in the DR21 filament which contains as much as 30~\% of the total mass in dense gas imaged by Motte et al. (2007) (see also \citealp{Schneider_prep}), 
and contains 
three 0.1~pc-scale MDCs: CygX-N44, CygX-N48, and CygX-N38. CygX-N44 hosts several spots of OH, H$_2$O maser emission \citep{PM90, LW97} and at high angular-resolution two strong peaks of continuum emission were identified \citep{Woody89, MWM91} (MM1 and MM2 in Fig~\ref{fig:intro_n2h}). The brightest peak, MM1, contains a "hot-core" (T$>$100~K) and shows cm continuum emission  \citep{MWM91} suggesting the recent birth of high-mass star(s). 
CygX-N48 and CygX-N38 have no mid-IR emission and host cold gas (T$<$40~K) \citep{MWM92}.

With such a high mass, the \object{DR21(OH)} clump qualifies as the best target in nearby regions (at less than 3 kpc)\footnote{We adopt a distance of 1.7 kpc~\citep{Schneider06}.} to probe the initial conditions of a rich and massive protocluster. It is expected to form a cluster as rich or even richer than the \object{Orion Nebula Cluster}\footnote{Assuming a star formation efficiency of 30~\%, the $\sim$7000~{\msol} of DR21(OH) would produce a stellar mass of 2100~{\msol} which is more than twice the stellar masses in the Orion Nebula Cluster~\citep{hillenbrand1997}.}.  We mapped DR21(OH) at high-angular resolution at 3~mm continuum, and in the {\nh} (J=1$-$0), and {\ch} (J=5$-$4) molecular lines with the PdBI to trace the cold, dense gas and warm gas with complex chemistry, respectively. 

\section{Observation and data reduction}
A mosaic with 3 pointings was obtained with the IRAM\footnote{IRAM is supported by INSU/CNRS (France), MPG (Germany) and IGN (Spain).} Plateau de Bure Interferometer (PdBI) with 6 antennas in Bq and Cq configurations between December 2008 and March 2009. Baselines range from 24 to 452 meters and T$_{\rm sys}$ was $\sim$100-150 and 60-80~Kelvins for the B and C tracks, respectively. The signal was correlated at 93.17613 and 91.98705 GHz in order to obtain the {\nh} (J=1--0) and {\ch} (J=5--4, K=0--4) lines respectively, with a velocity resolution of $\sim$0.25 {\kms}. Line-free continuum emission was obtained with 6 broad band correlator units. We used as phase calibrator the bright nearby quasar 2013+370 and as flux calibrator the bright evolved star MWC~349. 

We used the GILDAS software\footnote{See http://www.iram.fr/IRAMFR/GILDAS/} for the data reduction and analysis and applied the same method for all datasets. Zero-spacing information was available only for the {\nh} (J=1--0) transition \citep{Schneider_prep} and was combined with the interferometric data following standard procedures in GILDAS. \footnote{We exploit here  both the combined and the interferometric-only dataset of the {\nh} data, because the latter one filters out extended emission from the clump and shows the compact emission, which is the main focus of this paper.} The image was restored with a natural beam weighting in order to favor sensitivity. Clean components were searched for within a polygon, which excluded the borders of the mosaic with a higher noise level. The cleaning procedure was done with the Hogbom algorithm and components were searched down to 3$\sigma$ per channel noise level. The resulting parameters are shown in Table~\ref{tab:obsparam}. As a default in mosaic mode, the resulting clean maps are corrected for beam attenuation.

\section{{\nh} as tracer of the cold, young gas}

{\nh} is a tracer of cold, dense gas \citep{tafalla2002}. Since it is quickly destroyed in the heated envelopes of protostars, it shall represent the starless/unprocessed gas, which has either not yet fragmented or originates from recently formed pre/protostellar cores.

In a quasi-static view of star formation, the pre-collapse core traced by {\nh} is expected to evolve to a higher level of concentration by a progressive loss of support until it gets gravitationally unstable and collapses. In nearby low-mass star-forming regions such as $\rho$~Ophiuchi, {\nh} is indeed found to show single, narrow lines which are consistent with being close to or under gravitational collapse with only a small contribution from non-thermal motions \citep[][]{andre07}. 

\subsection{Compact {\nh} emission is concentrated in the MDCs}

In Fig.~\ref{fig:intro_n2h} a the spatial distribution of  {\nh} at low (IRAM 30m) and high (PdBI) spatial resolution is compared with the dust continuum emission which is taken as a reference to trace the total column density. 

Seen with a single-dish telescope the distribution of {\nh} differs from the dust continuum (Fig.~\ref{fig:intro_n2h}{\sl a}), as {\nh} peaks in the south of the clump between CygX-N48 and CygX-N38. This prominent north-south inhomogeneity indicates that generally the southern part of the clump is younger than the northern part. At high spatial resolution with the PdBI (Fig.~\ref{fig:intro_n2h}{\sl b}), the compact emission of {\nh} is clearly concentrated in the 3 known MDCs, CygX-N48, CygX-N38, and CygX-N44. This discrepancy of the distribution of {\nh} seen with the single-dish telescope compared to the interferometer shows that a large fraction of the {\nh} emission originates from extended structures that are filtered out by the interferometer. It also shows that the most compact components of {\nh}, and therefore the densest parts of this unprocessed, cold gas are associated with the MDCs in the clump. 

At high angular-resolution this spatial coincidence between the dust and {\nh} and the fact that both tracers seem to form structures at similar size scales of $\sim$0.1~pc further confirms the existence of a typical size-scale for MDCs and indicates that the MDCs are the main sites of present star-formation. In contrast, the hot-core region of DR21(OH)-MM1 is devoid of {\nh} suggesting that star formation there is in a later stage. A new, rather compact {\nh} core is seen north of CygX-N38 (Fig.~\ref{fig:intro_n2h}{\sl b}). This new structure coincides with very weak ($\sim$1.3~mJy/beam peak intensity) continuum emission and could either be a fluctuation in the global diffuse {\nh} gas, or could correspond to a new MDC in formation.

\subsection{Bulk motions in the {\nh} gas associated with the MDCs}
\label{sect:flows}

\begin{figure*}[!htpb]
\centering
\includegraphics[width=0.50\linewidth,angle=-90]{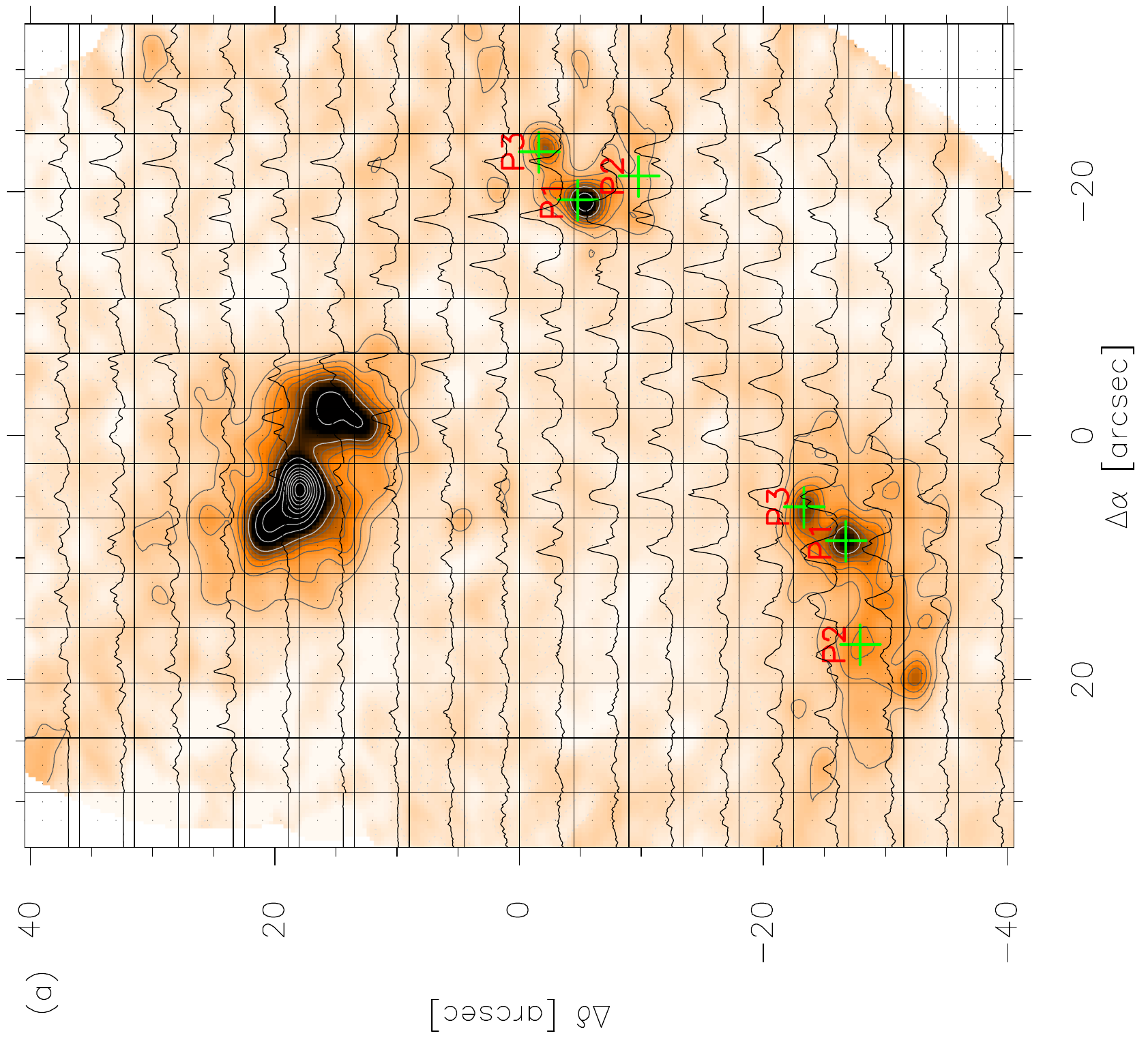}
\includegraphics[width=0.33\linewidth, angle=-90]{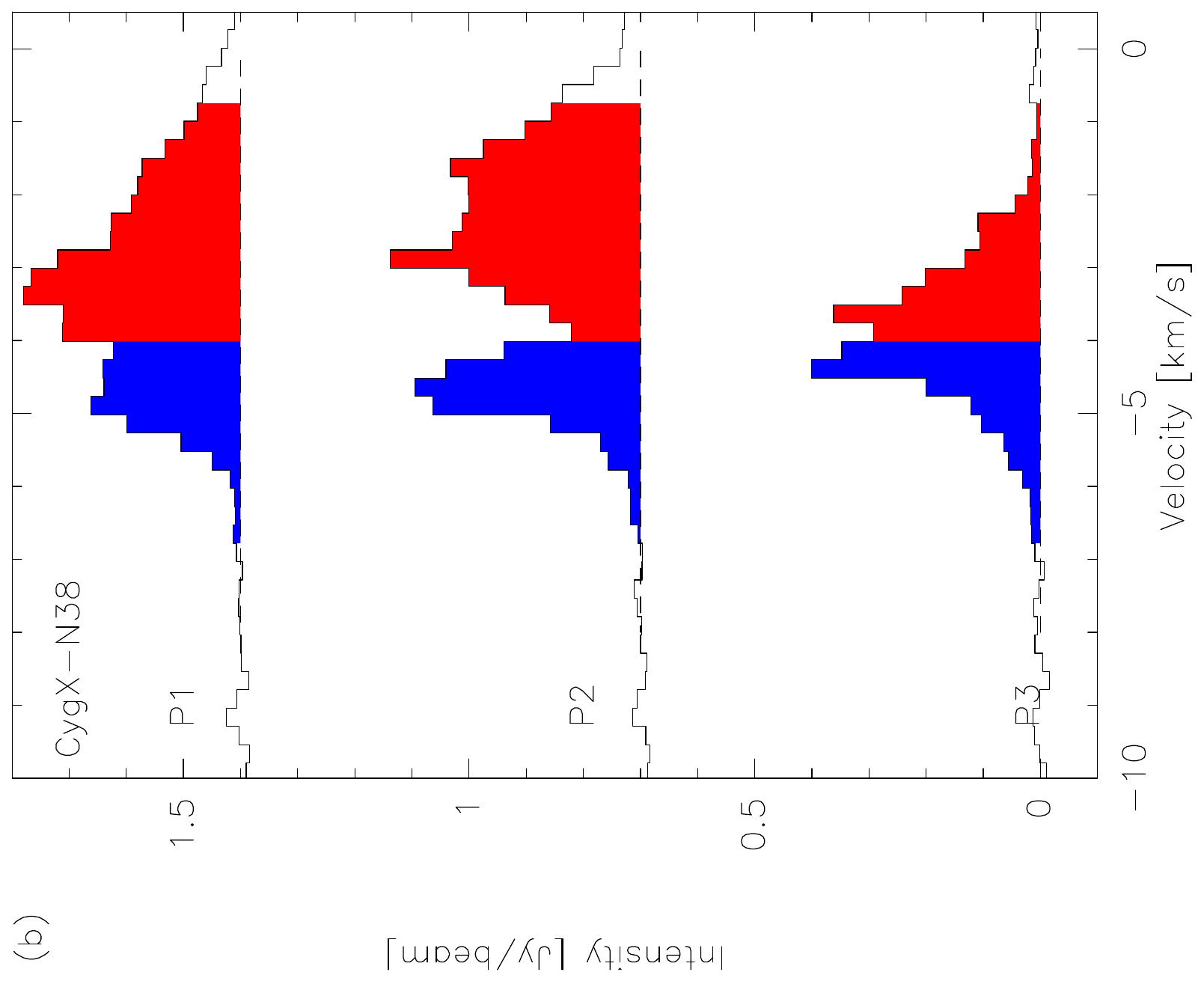}
\includegraphics[width=0.33\linewidth, angle=-90]{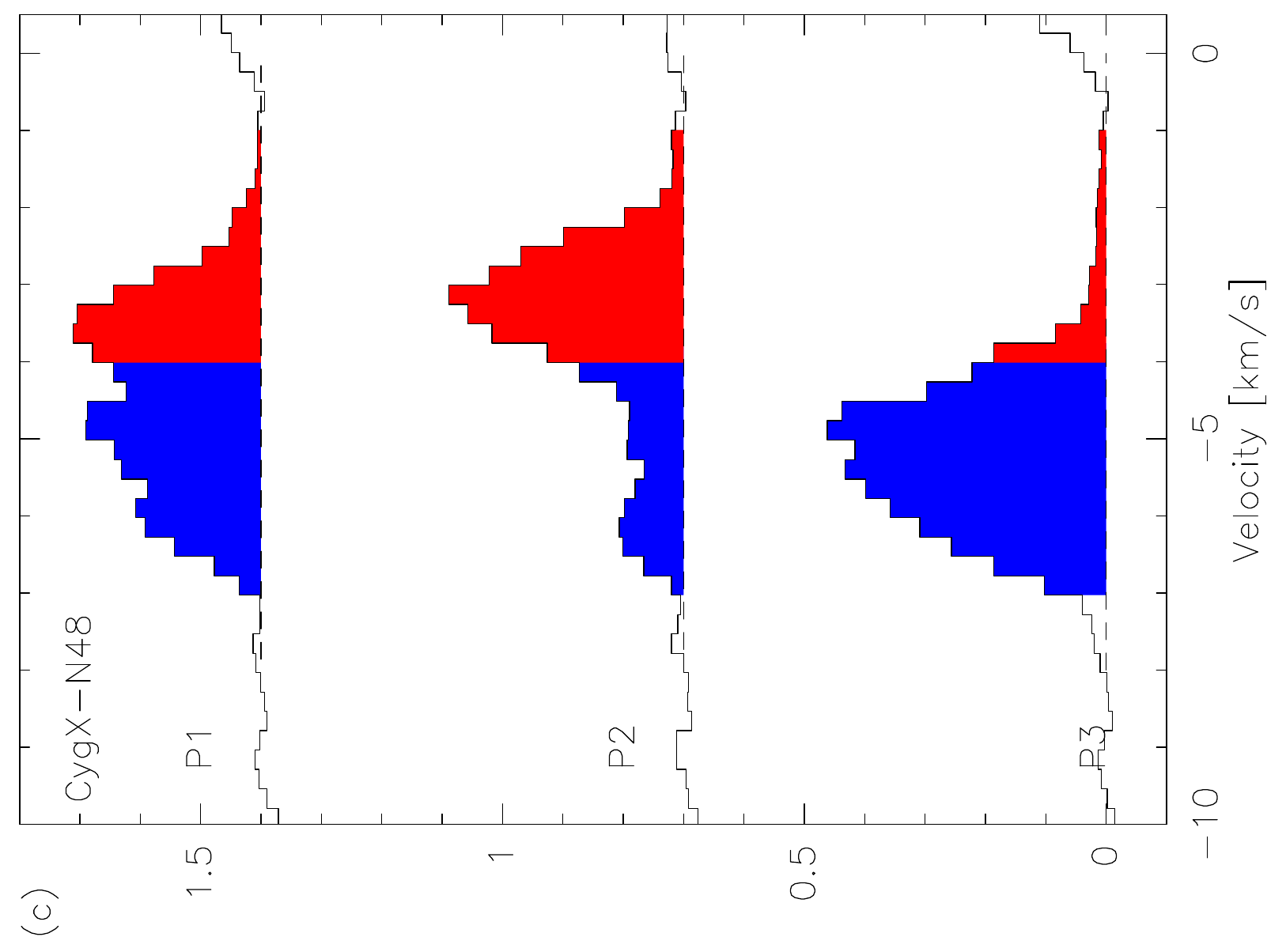}
\caption{{\bf a)} The map shows the 3mm continuum emission obtained with the PdBI (shown also in Fig.~\ref{fig:intro_n2h}). The overlayed spectra correspond to the isolated, F = 101--012 hyperfine component of the {\nh} (J=1--0) emission (PdBI data combined with zero-spacings from the IRAM 30m telescope) resampled on a 4{\arcsec} grid. The range of the $x$ velocity axis is from $-8$ to $+0.5$~{\kms}. The intensity range, y axis, scales from $-0.1$ to 0.4 Jy/beam. Dots mark the position of a grid sampled on half of the beam. Green crosses point the positions where spectra are extracted and shown in panel {\sl b} and {\sl c}. {\bf b)} and {\bf c)} Extracted spectra towards the 3 positions in CygX-N48 and CygX-N38. Red and blue shading indicates the integration ranges shown in Fig.~\ref{fig:vel_n2h_n48}. (For a color figure see the online edition.) }
\label{fig:online-spec}
\end{figure*}

At low angular-resolution with the IRAM 30m telescope the spectra show a single component \citep{Schneider_prep}, such as in $\rho$~Ophiuchi, but with a large velocity dispersion of $\sim$1.3~{\kms}. But while in $\rho$~Ophiuchi the line stays single at small scales~\citep{diFrancesco04}, the {\nh} emission seen with the interferometer clearly splits here into individual velocity components (Fig.~\ref{fig:online-spec}). 

This is a striking feature and an important difference to low-mass star-forming regions. Since we could include the zero-spacing information from the
IRAM 30m telescope, we are confident that there are no strong side-lobes and filtering which could modify the line profiles. Moreover, it seems that it is towards the MDCs that the
individual velocity components can be
recognized the best (see Fig.~\ref{fig:online-spec}a).
 
We show in Fig.~\ref{fig:vel_n2h_n48} the maps of integrated intensity of the
two velocity components which are associated with the two coldest MDCs, as well as position-velocity cuts through and across the intersection regions between the two velocity components. These cuts show velocity jumps indicated by arrows which have relative velocity difference of $2-3$~{\kms} and are referred
below as velocity shears. We note that the intersection regions of the velocity shears are close to the strongest peaks of continuum emission, but do not coincide exactly with them. 
The velocity shears are found offseted from the strong continuum sources. 
These {\nh} velocity patterns are very similar to what we have recently obtained using H$^{13}$CO$^+$ in isolated MDCs in Cygnus X~\citep{Csengeri10}. Similarly, we propose that the individual velocity components seen in {\nh} are also best interpreted as flows converging to the gravitational well of MDCs. These flows may carry a significant amount of angular momentum which could lead to rotational motions. 
For a 200~{\msol} core the Keplerian velocity at 0.1~pc is $\sim$2.9~{\kms} which is at a similar order as the gradient observed in velocity field. On the other hand the pv-cuts in Fig.~\ref{fig:vel_n2h_n48} show that the maximum velocities are not found at large offsets from the velocity jumps
similarly to what is expected in a Keplerian rotation, but they show a more complex distribution. The observed velocity pattern of the gas here is thus not compatible with a pure rotation of the cores. Furthermore an homogeneous sphere in rotation would also not show separated individual velocity components. 


It is surprising to see these flows in {\nh}, since it normally traces only pre-stellar cores which are gravitationally bound structures. On the other hand, it may just indicate that the pre-stellar gas is already dense and cold in the convergent flows which then form the even higher density structures at the center of the MDCs. 
Furthermore we note that the observed range of velocity (between $-6.5$ and +0.5~{\kms} over the whole clump) is larger than in \citet{Csengeri10} most probably because of the larger mass and gravitational well of the DR21(OH) clump.

\subsection{On the origin of the clump fragmentation into MDCs}

If the  $\sim$7000~{\msol} DR21(OH) clump would be governed only by thermal motions and gravity, it would fragment into a large population of $\sim $0.4~{\msol} fragments corresponding to the local Jeans-mass\footnote{The Jeans-mass is calculated here for $n\sim1.3\times$10$^5$~cm$^{-3}$, and $T=20$~K~\citep{M07}} which should concentrate in the central regions close to the gravitational center of the clump. 
Instead, the distribution of dense gas traced by {\nh} mostly reveals three centers of collapse, the MDCs. It is then not clear why the large scale, gravitationally driven flows split to predominately form a few MDCs dispersed over the clump. On the other hand, the observed hierarchical fragmentation is very similar to what is obtained in numerical simulations (e.g.\,\citealp{hennebelle2011}). These models indicate that fragmentation is provoked by a complex interplay between gravity, rotation, thermal/turbulent pressure and magnetic forces (\citealp{VS07}) which may make it difficult to clearly pinpoint the physical origin of the observed fragmentation properties.

\subsection{Are MDCs living for more than a crossing time?}

Since the gas dynamics dominate the evolution of the MDCs, the crossing times\footnote{Since the crossing-times are almost equal to the local free-fall times which are $\sim5\times10^4$~yr, it re-inforces the view by which self-gravity is the main driver of the observed dynamics (the whole DR21 filament is in global collapse).} of the MDCs, ranging from 5 to  $7\times10^4$~yr, should measure the life-time of the presently observed dense gas. On the other hand the large scale flows observed by \citet{Schneider_prep} are massive enough to continuously replenish the mass of the MDCs keeping them 'active' for a longer period if the local gravitational wells are spatially stable. The presence of still active {\nh} convergent flows at the location of each MDC favors this view of stable gravitational wells over time and thus suggests that the MDCs are re-filled (by large-scale flows) over longer times.

It is also striking to see that all three MDCs are situated close to embedded IR-bright sources likely excited by embedded massive stars (see Fig.~\ref{fig:online-spitzer}). This reveals the presence of a population of young stars which 
could have been formed at earlier time inside the same three MDCs.

\hspace{1cm}

\section{{\ch} reveals extended warm gas in each MDCs}

\subsection{Intriguing extended {\ch} emission in all MDCs}

A map of integrated {\ch} emission over all K=0-4 components is shown in Fig.~\ref{fig:intro_n2h}~d. {\ch} has long been considered to be produced only in hot cores and hot corinos  \citep[e.g.][]{kurtz2000, bottinelli2004} as a second generation molecule formed in the gas phase after ice sublimation resulting in very compact emission towards protostars. 
As expected, a bright, compact peak of {\ch} emission is centered on the hot core source MM1~(Fig.~\ref{fig:intro_n2h} d). 
Among the other continuum sources, we do not detect any similar strong and compact {\ch} emission which would indicate hot cores. 

\begin{figure*}[htpb]
 \begin{minipage}[b]{0.95\linewidth}
 \centering
   \includegraphics[width=0.305\linewidth]{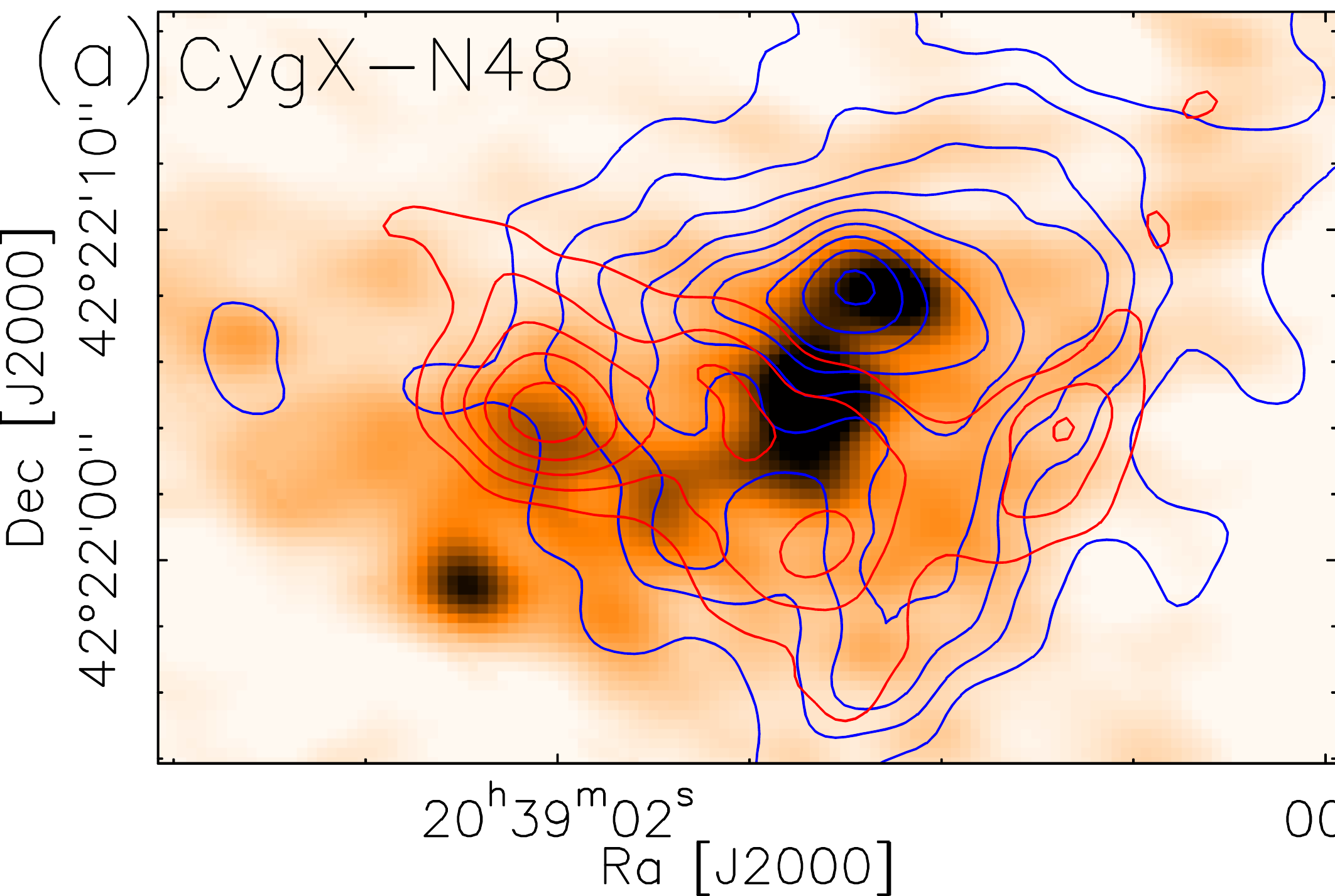}
   \includegraphics[width=0.3\linewidth]{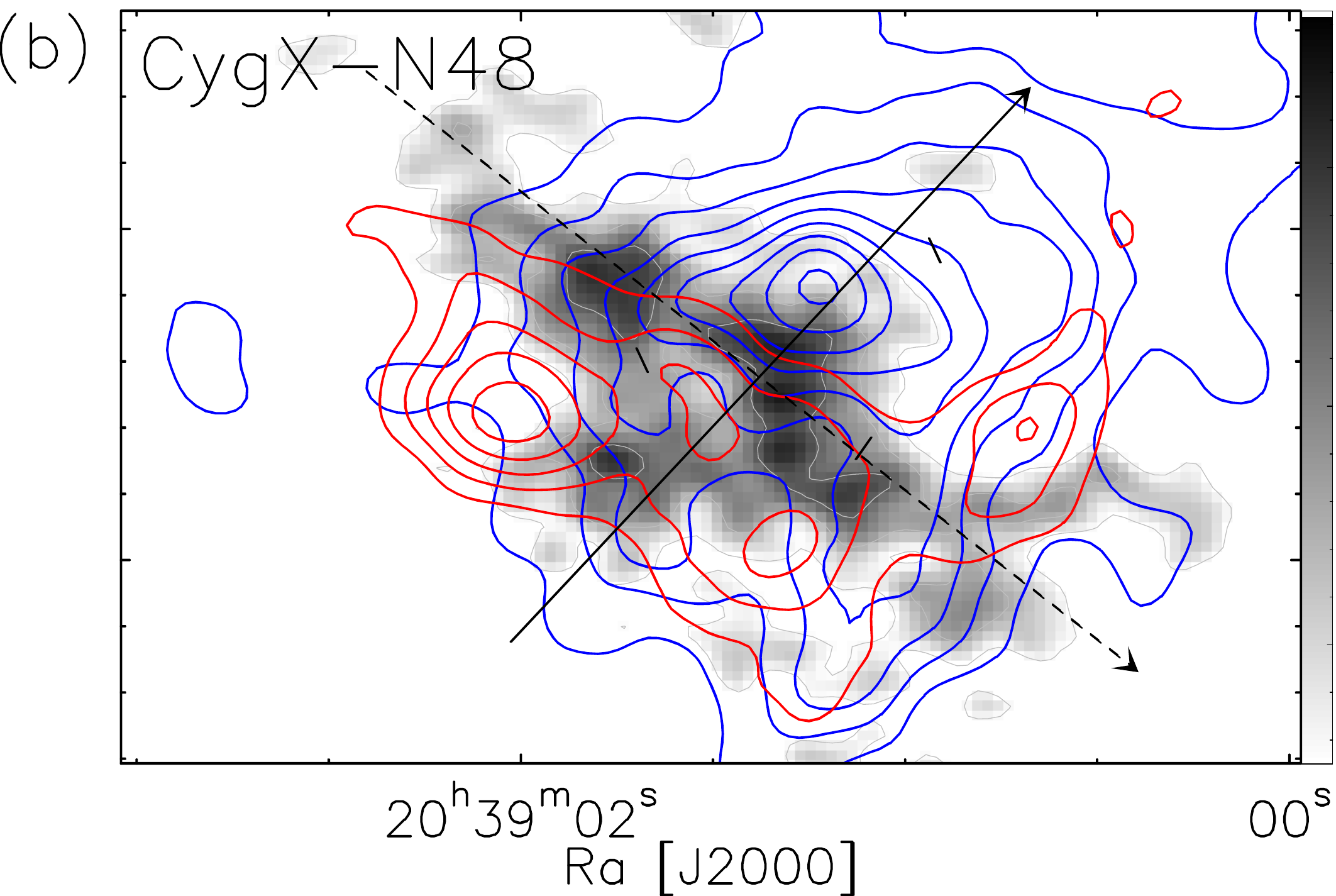}
    \includegraphics[width=0.35\linewidth]{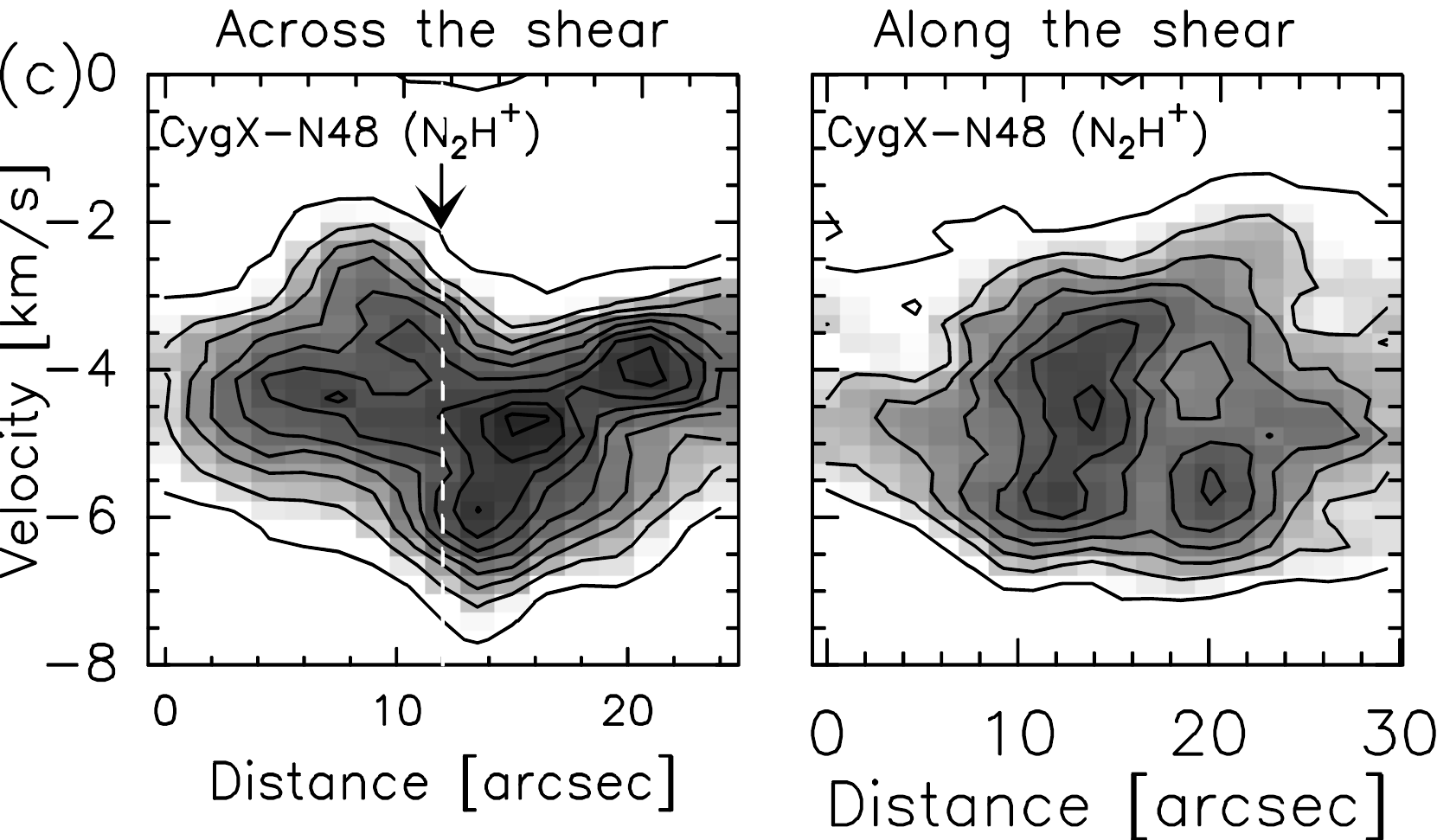}
  \end{minipage}
\begin{minipage}[b]{0.95\linewidth}
 \centering
  \includegraphics[width=0.31\linewidth]{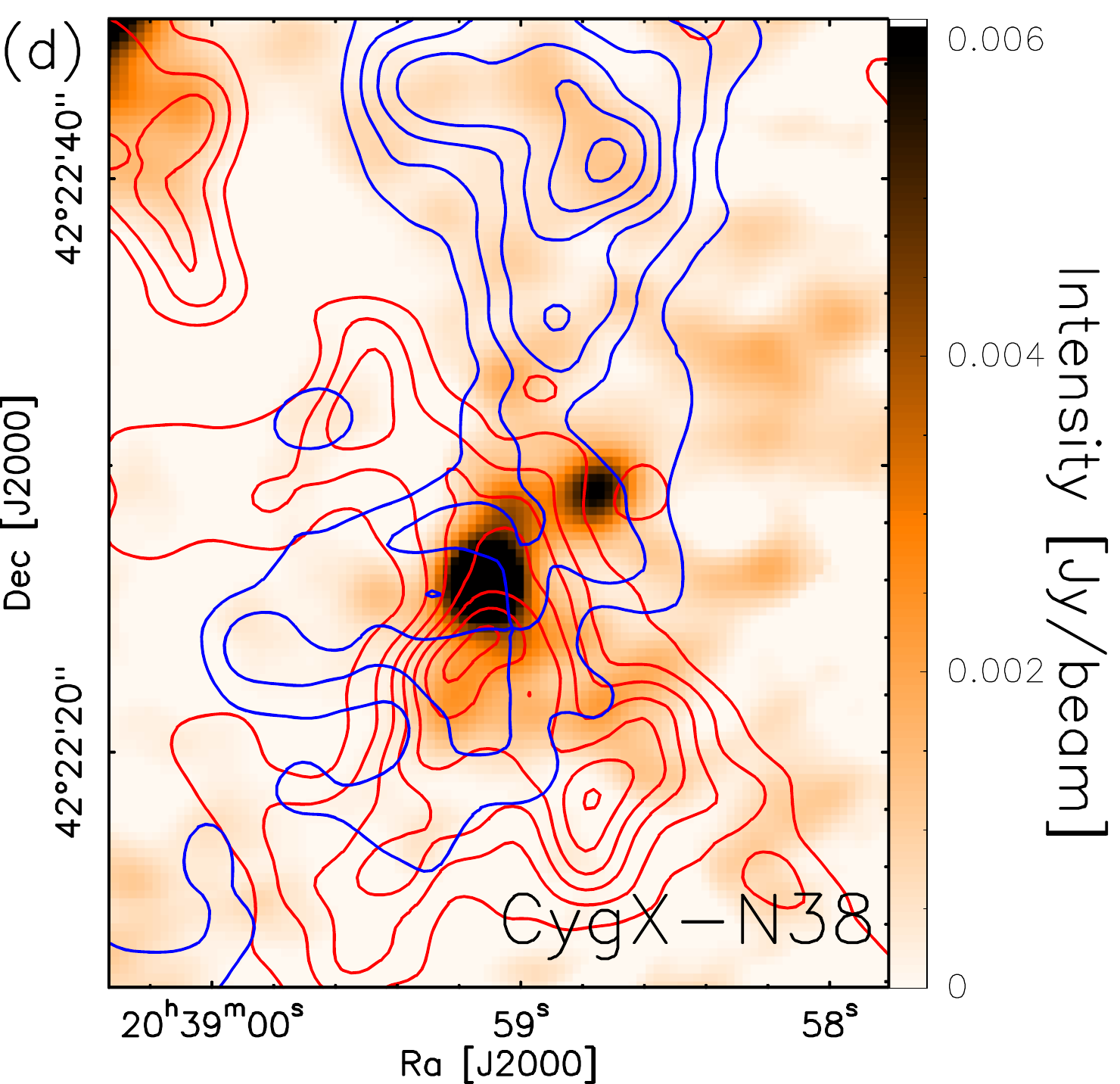}
   \includegraphics[width=0.31\linewidth]{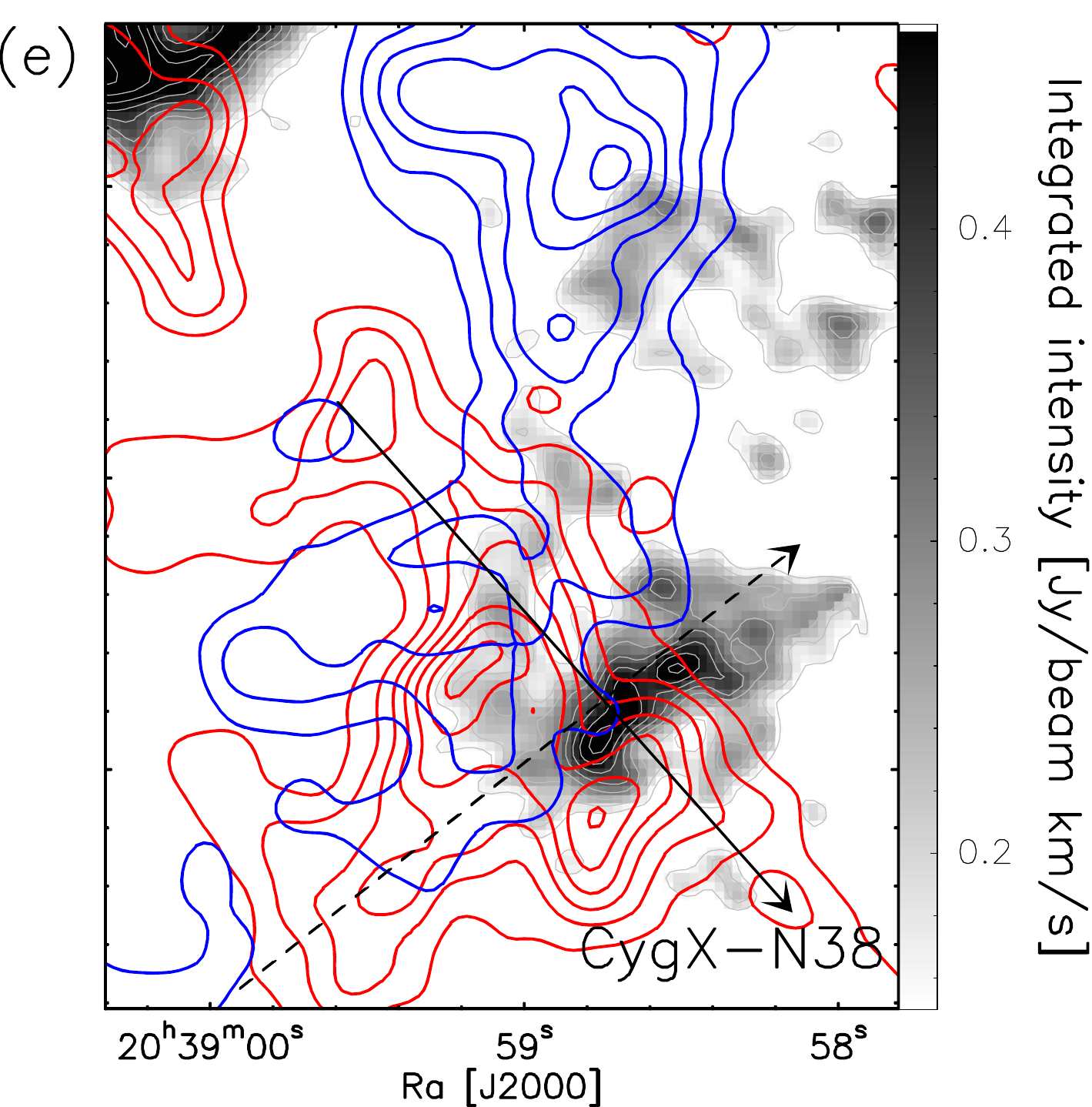}
    \includegraphics[width=0.35\linewidth]{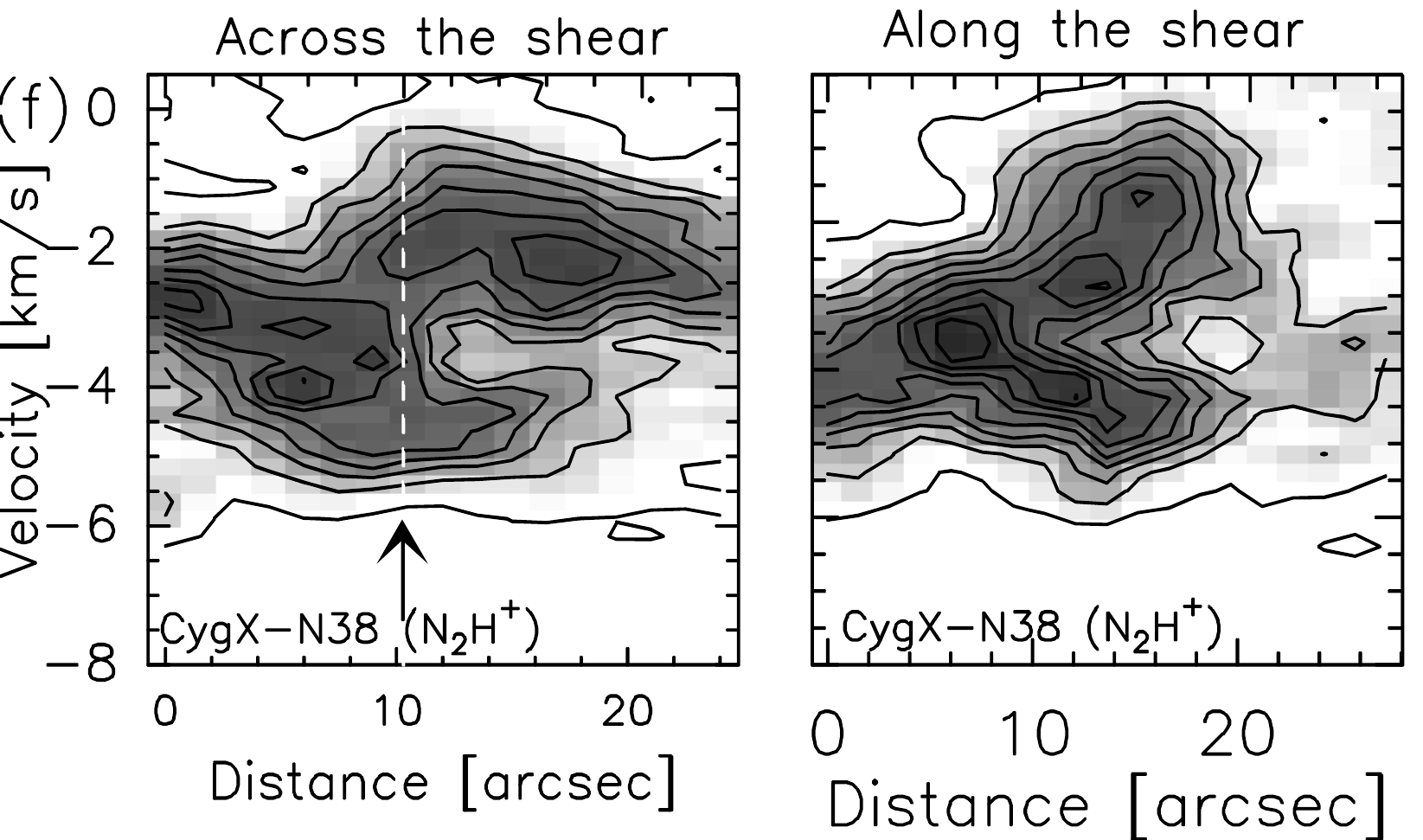}
  \end{minipage}
   \caption{{\bf a)} Zoom into the CygX-N48 MDC. The map shows the 3mm PdBI continuum emission. Contours display the {\nh} emission integrated between $-1.1$ to $-3.9$~{\kms} (red) and $-3.9$ to $-6.9$~{\kms} (blue) and go from 10$\sigma$ increasing by 10$\sigma$. {\bf b)} same as in (a) with the map showing the integrated emission over all K-components of {\ch}. {\bf c)} Position-velocity cuts of the {\nh} emission across (solid line) and along the shears (dashed line). {\bf d-f)} Same as (a-c) for the CygX-N38 MDC with {\nh} emission integrated between $-3.9$ to $-0.9$~{\kms} (red) and $-6.7$ to $-3.9$~{\kms} (blue) for the contours.} 
         \label{fig:vel_n2h_n48}
   \end{figure*}

In addition to the compact emission of MM1, widespread extended emission of {\ch} is  detected 
towards all MDCs of the region, including the colder ones, CygX-N38, CygX-N48. 
This extended emission is associated with the MDCs but is in detail almost entirely anti-correlated with the dust continuum (Fig.~\ref{fig:vel_n2h_n48}{\sl b} and {\sl e}). In CygX-N48, it forms an elongated structure almost perpendicular to the dust emission. In CygX-N38, it is shifted by 5{\arcsec} from the dust continuum. This indicates that {\ch} has to be over-abundant in the gas (in comparison with the bulk of gas traced by the continuum) to explain this particular spatial distribution. 

\subsection{Is {\ch} tracing shocks of the convergent flows?}

In Fig.~\ref{fig:vel_n2h_n48}{\sl b} and {\sl e}, the two main {\nh} velocity components are overlaid on the  {\ch} emission.  {\ch} seems to be bright at the location of the velocity shears (interfaces between the red and blue velocity components) discussed in Sect.\,\ref{sect:flows}, and which we interpret as tracing {\nh} convergent flows. It is striking that {\ch} coincides much better with these velocity shears  than with the dust continuum. Therefore we suggest that {\ch} could trace warm gas associated with the shocks expected at the convergence of the flows. 

{\ch} has actually been detected towards a young protostellar outflow shock in L1157 \citep[][]{codella2009}. The abundance of {\ch} can have significant contribution from grain surface chemistry \citep[][]{garrod2008} producing {\ch} directly on ices. A higher {\ch} abundance could originate in warm gas from ice sublimation. The involved velocities are however here smaller than in outflows, 2 to 3~{\kms} in projection which converts to 3.7 to 5.5~{\kms} if corrected for an average projection angle\footnote{The true post-shock velocities could be higher if the flows accelerate in the gravitational potential of the MDCs.}. 
From Fig.~3 in \citet{kaufman&neufeld1996} we see that the resulting post-shock gas temperature could be close to 100~K for a magnetized C-shock. At this gas temperature, the grains stay however cold and ice sublimation is not expected~\citep{Draine1983}. Partial desorption has however been recently proposed to explain gas phase abundance of complex molecules \citep[][]{garrod2007, arce2008, oberg2009}. The higher gas temperature and sputtering of grain mantle in the shock may have led to an increased desorption of {\ch} which could explain the detectable abundance of {\ch} in the gas phase which is otherwise virtually not present in the cold, dense gas.

\section{Conclusions}

It is striking that {\nh}, a classical tracer of cold, dense gas, shows flows associated with the highest density regions forming new high-mass stars. These flows are showing velocity shears which are found to spatially coincide with intriguing extended {\ch} emission. 

The evolution of the MDCs is driven by supersonic flows which are able to continuously supply a replenishment of material thanks to large scale flows (see \citealp{Schneider_prep}) while the MDCs form stars. As a consequence, they maintain active star-forming processes for longer than their free-fall times and crossing times if the convergent point of flows is stable in space over time. 

The spatial coincidence between the {\nh} velocity shears and the intriguing extended emission in {\ch} indicates that we have probably detected for the first time the effect of low-velocity, but high density shocks expected in the case of widespread convergent flows building up new high density gas in the MDCs.

\begin{acknowledgements}
T. Csengeri acknowledges support from the FP6 Marie-Curie Research Training Network ÕConstellation: the origin of stellar massesÕ
(MRTN-CT-2006-035890). This work was also supported by the ANR (Agence Nationale pour la Recherche) project ÒPROBeSÓ, number ANR-08-BLAN-0241. 
\end{acknowledgements}



{\it Facilities:} \facility{IRAM PdBI}, \facility{Spitzer}.



\newpage
\begin{appendix} 

%
\begin{table}[htdp]
	\centering
		\caption{Parameters of the observations. For the molecular lines the rms noise is given for 0.25~{\kms} spectral resolution.}
	\begin{tabular}{lcccc}
	\hline
	\hline
		PdBI		& synthesized beam	& P.A.		&  	 rms 	\\
					& [{\arcsec}$\times${\arcsec  ]} & & [mJy/beam] \\
	\hline
	3~mm continuum   & 2.11 $\times$ 1.9 & 84$^{\circ}$ &  0.3 \\
	{\nh} (J=1--0)	              &2.09 $\times$ 1.89 & 84$^{\circ}$ &  8.9	\\
 	{\ch} (J=5--4, K=0-4)  & 2.13 $\times$ 1.92 & 83$^\circ$ &   9.1\\      
	\hline
	\end{tabular}
	\label{tab:obsparam}
\end{table}

\begin{figure}[!htpb]
\begin{center}
\includegraphics[width=0.7\linewidth]{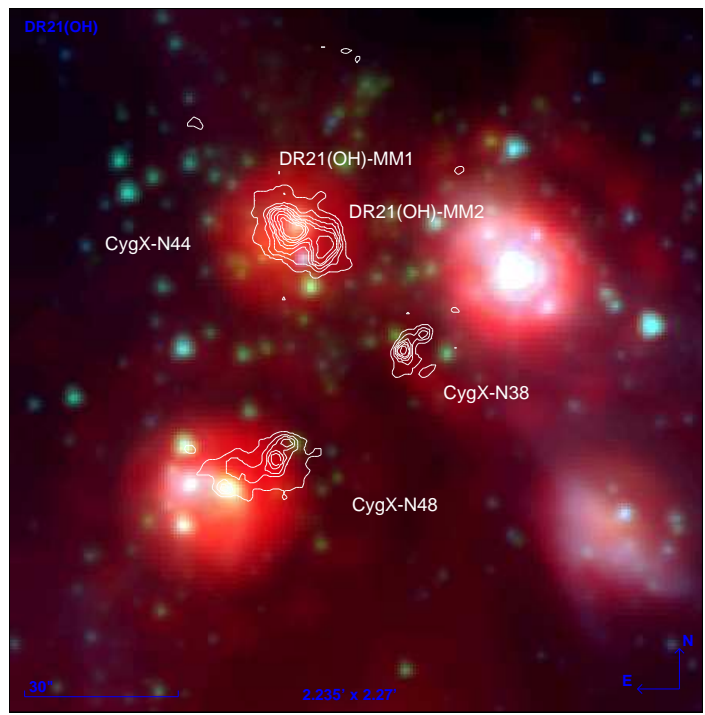}
\caption{Color composite image of Spitzer IRAC 3.6~$\mu$m, 4.5~$\mu$m and MIPS 24~$\mu$m~\citep{Hora}. Contours show the 3mm continuum emission obtained with the PdBI. Note the bright embedded sources close to the MDCs. Note also that there is a faint 24~$\mu$m emission at CygX-N38 which is an artifact of the MIPS PSF  due to the nearby bright source. (For a color figure see the online edition.) }
\end{center}
\label{fig:online-spitzer}
\end{figure}



\end{appendix}



\clearpage

\end{document}